\documentclass[a4paper,11pt]{article}
\usepackage{pos}

\title{Gluon dominance model and multiparticle production}

\author*[a,b]{E. S. Kokoulina}
\author[a]{A. Ya. Kutov}
\author[a]{V. A. Nikitin}
\author[c]{V. N. Riadovikov,}
\author[d,e]{R. G. Shulyakovsky}

\affiliation[a]{Joint Institute for Nuclear Research,\\
  Joliot-Curie str., 6 Dubna, Moscow region, Russia}

\affiliation[b]{Gomel State Technical University,\\
Prospect Octiabria, 48, Gomel, Belarus}

\affiliation[c]{Institute of High Energy Physics,\\
sqr. Science, 1, Protvino, Moscow region, Russia}

\affiliation[d]{Institute of Applied Physics of NAS of Belarus,\\
st. Akademicheskaya, 16, Minsk, Belarus}

\affiliation[e]{Institute for Nuclear Problems, BSU,\\ st. Bobruiskaya, 11, Minsk, Belarus}

\emailAdd{kokoulinar@jinr.ru}
\emailAdd{akutov@rambler.ru}
\emailAdd{nikitin@jinr.ru}
\emailAdd{riadovikov@ihep.ru}
\emailAdd{shulyakovsky@iaph.bas-net.by}

\abstract{The gluon dominance model describes multiparticle production of secondary particles at high energies in lepton and hadron interactions, including annihilation processes and heavy quarkonium decays. According to this model, the multiparticle process is divided into two stages. The first stage describes the development of a quark-gluon cascade as a Markov branching process in the region of perturbation QCD. For the second stage, the transformation of quarks and gluons into observable hadrons (hadronization), a phenomenological scheme is proposed. It is universal and based on an experiment. The gluon dominance model demonstrates good agreement with data over a wide energy region.
It testifies that in hadron interactions valence quarks remain in the leading particles, and gluons are the sources of secondary hadrons. Quantitative estimates of the model parameters confirm the fragmentation mechanism of hadronization in leptonic interactions and the recombination mechanism in hadronic ones. The model description of the experimental distributions on the number of neutral pions in proton interactions at 50 GeV beams in the high multiplicity region are presented for the first time. It is shown that the main contribution to this region is made by gluon fission. These results can be useful in planning of future experiments.}

\FullConference{53rd International Symposium on Multiparticle Dynamics, September, 21-28 2025,
EISA, Corfu, Greece}

\begin{document}
\maketitle

\section{Introduction}
Multiparticle production (MP) of secondary particles accompanies any experiment at high energies. Its study began in cosmic rays and continues at modern accelerators in interactions of protons, heavy ions and at electron-positron colliders. Over the past 50 years, the energy of colliding particles has been increased to the TeV region. Physicists are developing new technologies that will increase energies to hundreds of TeV.

The number of secondary particles formed in a collision is called multiplicity.
A distinction is made between the multiplicities of charged, neutral particles, and the total multiplicity (their sum). The number of secondary particles produced in a single interaction varies randomly from one event to another. Their average multiplicity increases with energy approximately logarithmically.

The study of MP stimulated the development of phenomenological models, including the theory of strong interactions, quantum chromodynamics (QCD) \cite{QCD}. 
QCD allows one to calculate the characteristics in the region of large momentum transfers, where the constant of strong interaction is small and perturbation theory (PT) \cite{QCD1} is applicable. QCD speaks the language of quarks and gluons, which are not clearly observed in experiments. The transition from quarks and gluons to observable hadrons in this region of QCD is difficult.

To pass from quarks and gluons to experimentally observed hadrons, phenomenological models are developed \cite{Model1,Model2,Model3,GDM1,TSM,GDM2,GDM3,GDM4}, and rather complex time-consuming lattice calculations are performed. Before every experiment, physicists simulate the operation of their setup using Monte Carlo generators.

At the same time, it often turns out that the results of modeling differ significantly from the experimental ones. This forces us to make changes, or, as physicists say, to adjust the generators to data. And so at each increase of energy. A particularly large discrepancy is observed in the region of high multiplicity, which significantly exceeds its average value.

This work is based on the results of study of proton interactions with the 50 GeV beams at the U-70 accelerator at IHEP (Protvino) carried out by the SVD-2 Collaboration \cite{LHEP}. The main goal of this experiment was to search for the collective behavior of secondary particles in the high multiplicity region, both charged and neutral pions. Pions are the dominant particles in this region.

The main results of this experiment: topological cross-sections were measured with an advancement of three orders of magnitude down compared to previous measurements at the Mirabelle \cite{Mirab} setup, a signal of the formation of a pion (Bose-Einstein) condensate \cite{BE} was detected, the disappearance of the effect of leading secondary particles \cite{Leader} was shown, distributions of the number of neutral pions were reconstructed, and an increased yield of particles was detected in a narrow region of the polar angle, which we interpret as bremsstrahlung of gluons by quarks \cite{Leader}.

Before carrying out these studies, our Collaboration performed Monte Carlo simulations using the PYTHIA generator, and in parallel developed the gluon dominance model (GDM) \cite{GDM1,GDM2}. The simulation results showed that the topological cross sections are underestimated (by three orders of magnitude) at the maximum multiplicity $n_{ch}$ = 18, registered at the Mirabelle bubble chamber.

The gluon dominance model (GDM) was originally developed to study the MP in electron-positron annihilation processes \cite{GDM1,TSM}. This model is a convolution of two stages. The first stage, called the quark-gluon cascade, is described by the differential-difference equations \cite{KUV,Giov} as a Markov branching process. The cascade develops due to the fission of gluons $g\to gg$ and the bremsstrahlung of the gluon $q\to qg$ by the quark, described by PT QCD \cite{KUV,Giov}.

\begin{figure}[h]
	\leavevmode
	\centering
	\includegraphics[angle=0, width=0.45\textwidth]{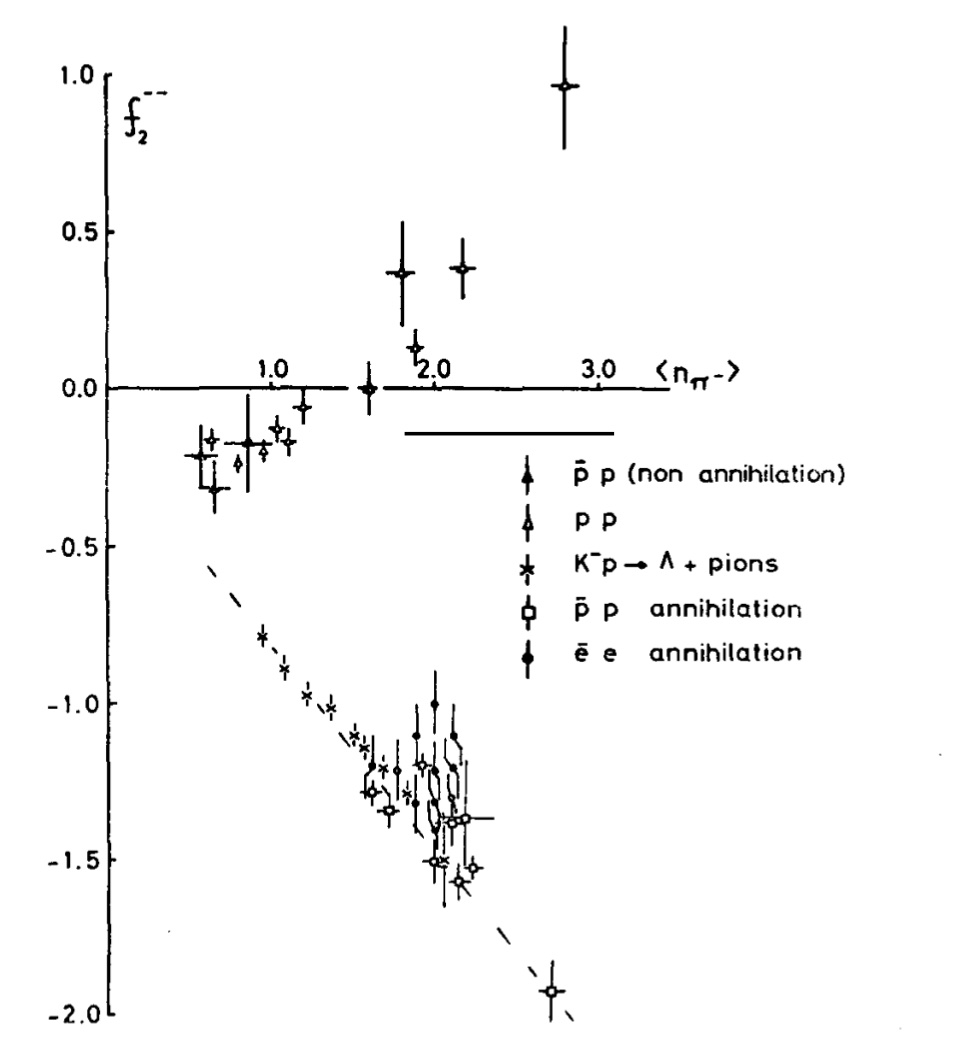}
	\includegraphics[angle=0, width=0.4\textwidth]{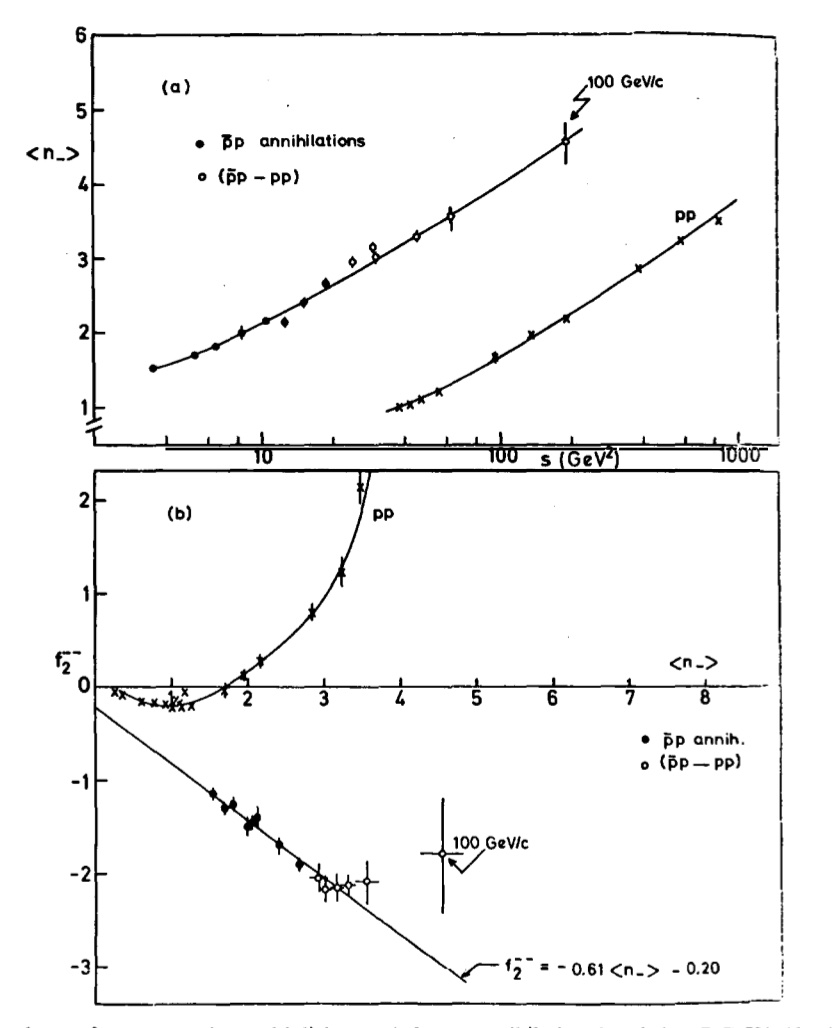}
	\caption{Experimental values of $f_2$ as a function of the mean multiplicity of negative pions $< n_{\pi _{-}}>$ for annihilation and non-annihilation processes \cite{Rush}.} 
	\label{fig1}
\end{figure}

To describe the second stage, hadronization, a phenomenological scheme is used. Its choice is based on the experimental behavior of the second correlation moment \cite{Rush}
\begin{equation}
	\label{eq}
	f_2 = D_2 -\overline n = \overline {n(n-1)} -\overline n^2,
\end{equation}
where $\overline n$ and $D_2$ are an average multiplicity and variance, the upper bar means averaging. $f_2$ changes sign from minus at low energies to plus at high. In addition, the sizes of the region of negative values $f_2$ differ significantly between proton collisions and annihilation processes as it shown in Figure \ref{fig1} for the average multiplicity of negative pions, $<n_{\pi_{-}}>$ for $e^+e^-$, $p\overline p$, $K^-p$, and $pp$ (left) and for $p\overline p$ and $pp$ (right).

The second correlation moment calculated for the multiplicity distributions (MD) in the quark jets is always positive at any energy. It takes negative value for the gluon jets unless the average number of newborn gluons, $\overline m$,  exceeds 1. Considering that at low energies the $qg$-cascade is insufficiently developed and the hadronization stage predominates, we choose for it the binomial (Bernoulli) distribution with negative $f_2$ as the most suitable
\begin{equation}
	\label{eq1}
	P_n^h = C_{N_p}^n \left ( \frac{\overline n_p^h}{N_p}\right )\left(1- \frac{\overline n_p^h}{N_p} \right )^{N_p-n},
\end{equation}
where index $p$ corresponds to $q$ (quark) or $g$ (gluon), $C_{N_p}^n $ is the binomial coefficient, $\overline n_p^h$ and $N_p$ determine the average and maximum possible number of hadrons formed from one parton $p$ at the hadronization stage.

The scheme combining two stages of MP was originally called the two-stage model 
(TSM) \cite{GDM1}. The MDs obtained in it agree well with the data on electron-positron annihilation up to $\sim$ 200 GeV \cite{GDM1,GDM3}. Moreover, the average number of hadrons $\overline n^h$ formed from one gluon at the hadronization stage remained approximately constant and close to 1, which is in good agreement with the hypothesis of local parton-hadron duality (LoPAD) and the fragmentation nature of hadronization \cite{Muel}.

To predict the behavior of topological cross sections in proton collisions before carrying out our experiment at the SVD-2 facility located at the U-70 accelerator at IHEP, the TSM was used. It only needed to be modified for proton-proton collisions.
Including all valence quarks into $qg$-cascade showed that the hadronization gluon parameter $\overline n^h$ becomes significantly less than 1. it contradicted to $e^+e^-$-annihilation results. 

A consistent decrease of the valence quarks was leading to the growth  hadronization parameter. And only with their complete exclusion $\overline n^h$ did  approach 1, slightly exceeding it, in contrast to $e^+e^-$-annihilation. We explained this by the change of hadronization mechanism in the quark-gluon medium to recombination, in contrast to fragmentation in vacuum, as proposed by the head of the theoretical department of BNL B.~Muller \cite{Muel}.

Exclusion of valence quarks from the model means their conservation in the leading particles. At the same time, the birth of secondaries occurs due to energetic gluons, which we called active. Subsequently, TSM was renamed the gluon dominance model (GDM) \cite{GDM4}.

This model is successfully applied to describe the MD in a three-gluon decay of heavy quarkonia \cite{Tomsk} and in proton-antiproton annihilation \cite{NPCS24}.

At present, JINR is building the NICA collider, where heavy ions, polarized protons and deuterons will be accelerated. Ion collisions will be performed at the MPD facility, and polarized particles at the SPD. Our group proposes to include in its physics program the study of MP in proton collisions, started at IHEP at the SVD-2 facility, which will be able to register rare events with high multiplicity.

The phenomenological model, GDM will contribute to a more profound study of the gluon structure of the nucleon. One of the important physical problems included in the SPD program is the study of the gluon component of the proton and neutron (in the deuteron) \cite{TDR}. It should be noted that the physics program planned for the future EIC (Electron-Ion Collider) at BNL is also aimed at clarifying the role of gluons in the formation of visible baryon matter in the Universe \cite{eRHIC}.

In this paper we present the main results on the study of MP in lepton and hadron interactions obtained within the framework of GDM and at the U-70 accelerator. In the second chapter we calculate the correlative moment $f_2$ in $e^+e^-$ annihilation. The third chapter contains information on the experimental results obtained in $pp$ collisions and their description in GDM.

In the fourth chapter, an expression for describing the MD of neutral pions with and without gluon fission is constructed within the framework of GDM. We managed to register events with a large total multiplicity (36 pions), where more than half of the kinetic energy ($\sim 64 \%$) is transformed into mass. It is shown that the tail of high multiplicity of $\pi ^0$ mesons is due to fission gluons, even if their number is small.

MDs in the region of small total multiplicity exhibit oscillatory (hopping) behavior due to the charge conservation law (even number $n_{ch}$). In distributions over odd and even numbers of pions, these oscillations disappear, returning to a smooth curve.

\section{GDM and electron-positron annihilation}
The GDM has been proposed to describe the MD of secondary hadrons in $e^+e^-$-annihilation. By that time, QCD already existed, and experimental MDs were available in a wide energy range. In the area of applicability QCD (PT), differential-difference equations for generating functions (GF) were constructed, corresponding to the MD in quark and gluon jets \cite{KUV,Giov}.

By definition, the GF $Q(s,z)$ is a convolution of the MD
\begin{equation}
	\label{eq3}
	Q(s,z) = \sum _n {P_n z^n}, 
\end{equation}
where $s$ is the square of initial energy. $P_n(s)$ can be obtained from $Q(s,z)$
$P_n(s)$ with the auxiliary variable $z$
\begin{equation}
	\label{eq4}
	P_n = \frac{1}{n!}\frac{\partial ^n}{\partial z^n}Q(s,z)|_{z=0}.
\end{equation}
It is also possible to calculate the correlation moments $F_k(s)$, which are in demand when studying multiple processes (MP)
\begin{equation}
	\label{eq5}
	F_k(s) = \overline {n(n-1)(n-2)\dots (n-k+1)} = \frac{\partial ^k}{\partial z^k}Q(s,z)|_{z=1} 
\end{equation} 
(the horizontal line means averaging over the number of particles).
The average multiplicity $\overline n$, the variance $D_2$ and the second correlation moment $f_2$ are determined through the GF:
\begin{equation}
	\label{eq6}
	\overline n = \sum _n n P_n = \frac {\partial }{\partial z}Q(s,z)|_{z=1},
\end{equation}
\begin{equation}
	\label{eq7}
	f_2 = \overline {n(n-1)} -{\overline n^2} = Q^{''}(s,z)|_{z=1} -( Q^{'}(s,z)|_{z=1})^2, D_2 = f_2 +\overline n.
\end{equation}

At the description of MP as a Markov branching one \cite{Giov}, the energy variable is taken as an evolutionary parameter. Two elementary events are taken into account: the quark bremsstrahlung ($q\to q+g$) and the gluon fission ($g\to g+g$). The probabilities of these events are determined in QCD (PT). The formation of a $q\overline q$-pair from a gluon at the $qg$-cascade stage is suppressed according to QCD estimates, so it is neglected.

The system of differential equations for the GF of quark and gluon jets allows us to determine the MD $P_m^P$ ($P = q, g$) \cite{Giov}. For the quark jet, this is the Polya-Egenberger distribution or negative binomial distribution (NBD)
\begin{equation}
	\label{eq8}
	P_m^q =\frac{k_p(k_p+1)\dots (k_p+m-1)}{m!}\left ( \frac {\overline m}{\overline m +k_p}\right )^m \left (\frac{k_p}{\overline m +k_p}\right )^{k_p},
\end{equation}
where the parameter $k_p$ is determined by the ratio of probabilities of two elementary events (quark bremsstrahlung to gluon fission), $\overline m$ is the average gluon multiplicity.
MD in gluon jets is the Farry distribution \cite{Giov}
\begin{equation}
	\label{eq9}
	P_m^g =\frac{1}{\overline m}\left ( 1-\frac{1}{\overline m}\right )^{m-1}.
\end{equation}

Energy of partons at their fission decreases up to a certain value. And then, the hadronization starts. At this stage, gluon can also continue to make fission (in the non PT QCD already) and some of them decay into $q\overline q$-pairs (we call these gluons active) forming secondary hadrons.

The TSM was applied to describe the MD in $e^+e^-$-annihilation. This process can be represented as the following sequence of events: the formation of a virtual photon or $Z^0$-boson, the birth of a $q\overline q$-pair, the development of a $qg$-cascade, and hadronization described by the MD (\ref{eq1}).

It is assumed that at the second stage there is no significant momentum transfer between partons, the so-called soft discolorating, which allowed the convolution of two stages to be performed and to obtain the hadron MD used to describe the \cite{GDM1,TSM} data.

The TSM showed good agreement with the experiment up to the maximum accessible energy ($\sim $ 200 GeV). At the same time, the average gluon multiplicity demonstrated a logarithmic increase with energy. The hadronization parameter for the gluon $\overline n^h$ turned out to be constant and close to unity over the entire studied interval (from 14 to 189 GeV) \cite{GDM3}.

This behavior corresponds to the fragmentation mechanism of hadronization in vacuum, when one gluon fragments into one hadron. It is also consistent with the hypothesis of local parton-hadron duality (LoPAD), proposed for the relationship between parton and hadron average multiplicities \cite{QCD1}.

The parameter $k_p$, describing the MD in a $qg$-cascade (\ref{eq8}), showed an excess over unity, which indicates the predominance of quark bremsstrahlung over their fission.

The second correlation moment, calculated from the convolution of two stages,
\begin{equation}
	\label{eq13}	
	f_2 = F_2 - F_1^2 = \left[\alpha ^2 \frac {\overline m ^2}{k_p}+\alpha ^2 \overline m - \frac {2+\alpha \overline m}{N}\right ](\overline n^h)^2.
\end{equation} 
takes negative values at low average gluon multiplicity $\overline m$ and becomes positive as it increases with energy. The parameter $\alpha $, equal to the ratio of hadronization parameters ($\overline n^h$) of the gluon to the quark, turned out to be less than one, which indicates that gluon jets are softer than quark jets in $e^+e^-$ annihilation at hadronization.

\section{GDM and hadron interactions}
The successful description of MD in $e^+e^-$-annihilation processes due to the phenomenological consideration of hadronization allowed the modification of GDM for studying hadron interactions. The need to develop this approach arose in connection with the beginning of searching for collective phenomena in proton interactions at high multiplicity at the U-70 accelerator at IHEP.

This project was called "Thermalization" \cite{LHEP}. Its participants were IHEP, JINR and SINP MSU. The experiment was carried out at the SVD-2 setup. It included a silicon vertex detector, a liquid hydrogen target, three modules of drift tubes, a magnetic spectrometer (19 proportional chambers inside a large magnet) and an electromagnetic calorimeter.

To suppress low-multiplicity events, a scintillation hodoscope was made, called a high-multiplicity trigger, with a lower threshold of 8 charged particles. An algorithm was included in its operation to exclude recording of events on the hodoscope itself. This trigger made it possible to measure topological cross-sections \cite{LHEP} three orders of magnitude lower than those previously obtained at the Mirabelle \cite{Mirab} facility.

The modification of TSM was carried out by comparing the model MD with the topological cross sections measured at the Mirabelle \cite{Mirab} setup. Initially, all three pairs of valence quarks and a certain number of gluons appearing at the beginning of collision were included in the scheme. Hadronization was taken into account in the same way as in $e^+e^-$-annihilation. The parameter $\overline n^h$ found when comparing this scheme with data on $pp$-collisions at 70 GeV/$c$ turned out to be significantly less than unity.

The natural step was to reduce the number of valence quark pairs in the scheme to two, then to one. But the value of the hadronization parameter remained, as before, less than one. And only when all valence quarks were excluded from the scheme, the parameter $\overline n^h$ overcame the value of 1 and slightly exceeded it. The explanation of changing in behavior of the gluon hadronization parameter was obtained after experiments at the Relativistic Heavy Ion Collider (RHIC) at Brookhaven National Laboratory (BNL) \cite{Muel}, as a transition to recombination mechanism of hadronization.

After modification, the TSM was renamed the gluon dominance model (GDM), since the sources of secondary hadrons in it are gluons. It was implemented in two schemes, each of which consisted of two stages. The first scheme took into account the fission of gluons arising at the moment of interaction. The second scheme omitted their fission. At the second stage, hadronization occurred, described by the same Bernoulli distribution (\ref{eq1}).

The expressions for MD of hadrons that has been used to describe the data in the first scheme are
$$
P_n(s) = \sum _{k=1}^{MK} \frac{\overline k ^k e^{-\overline k}}{k!}
\sum _{m=k}^{MG} \frac{1}{{\overline m}^k} \frac{k(k+1)(k+2)\dots (m-1)}{(m-k)!} \left ( 1- \frac{1}{\overline m }\right )^{m-k} \times
$$
\begin{equation}
	\label{eq16}	
	\times C_{\alpha mN}^{n-2} \left(\frac{{\overline n}^h}{N} \right )^{n-2} \left(1-\frac{{\overline n}^h}{N} \right )^{\alpha mN-(n-2)}
\end{equation}
and for second one
\begin{equation}
	\label{eq17}
	P_n(s) = \sum _{m=1}^{ME} \frac{\overline m ^m e^{-\overline m}}{m!}
	C_{mN}^{n-2} \left(\frac{{\overline n}^h}{N} \right )^{n-2}\left(1-\frac{{\overline n}^h}{N} \right )^{mN-(n-2)}.
\end{equation}

The expression (\ref{eq16}) is given for the first time. Here the summation variables $k$ and $m$ correspond to the number of gluons at the initial moment of impact (random appearance) and after their ending of their fission, respectively. Their upper bounds are determined from the the comparison with data. The difference $n-2$ takes into account the two initial protons. The parameter $\alpha $ determines the fraction of active gluons fragmenting into hadrons. At 70 GeV/c it is approximately 1/2. The number of fission gluons in GDM ($MG$) grows indefinitely as their energy decreases.

Both schemes agree quite well with the data. Determined parameters of hadronization in both cases are comparable, exceed unity and slowly increase with energy. We interpret this as a manifestation of the recombination mechanism of hadronization. The region of high charge multiplicity registered in the experiment at U-70 can be described only by the scheme with gluon fission \cite{GDM3}.

GDM estimated the fraction of events in which a proton is recharged into a neutron with charge exchange to a neutral pion. It is equal about half of all events. It is consistent well with the data \cite {Per}. This was achieved by moving into the region of high multiplicity \cite {Egle}.

\begin{figure}[ht] 
	\leavevmode
	\centering
	\includegraphics[angle=0, width=0.45\textwidth]{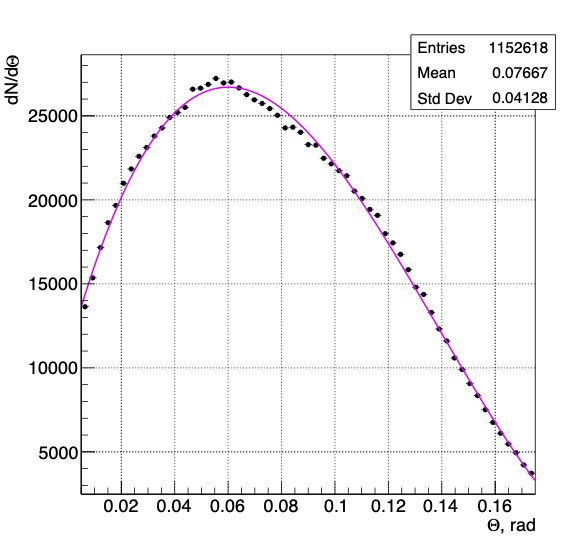}
	\includegraphics[angle=0, width=0.45\textwidth]{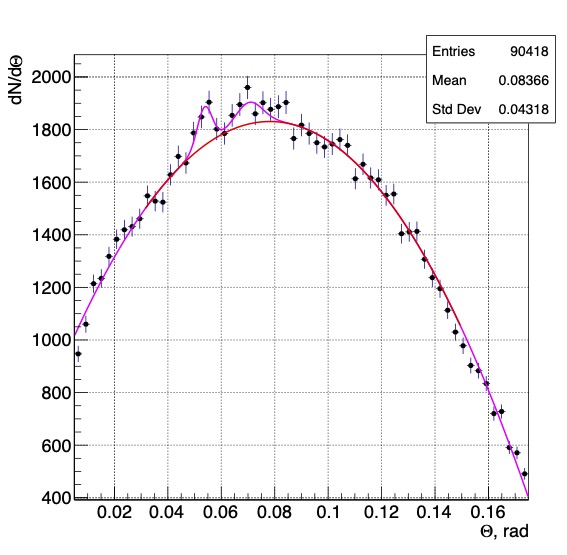}
	\caption{Polar angle distributions for $n_{ch}$ < 13, (left) and $n_{ch}$ > 13, (right). Fitting of two peaks is carried out by Gauss function, background -- by the polynomial of the third power. } 
	\label{fig17}
\end{figure}

We also associate the manifestation of gluon component with the discovered double-humped structure in the angular distribution by the polar angle as shown in Fig. \ref{fig17} in the region of high multiplicity ($n_{ch}$ > 13, right), which is absent at low ($n_{\textnormal {ch}}$ < 13, left).

Interpreting these peaks as a manifestation of Cherenkov radiation by quarks of gluons, we estimated the refractive index of the resulting nuclear medium. Its value differs from unity insignificantly, which indicates its transparency. The corresponding estimate obtained from the RHIC accelerator experiment is close to 3.

It was also found that with increasing multiplicity, the longitudinal component of the momentum decreases, approaching the transverse one, which remains practically constant \cite{Leader}. Such behavior of the hadron system indicates the disappearance effect of leading particles, and/or its isotropic decay.

GDM describes MD in processes of $p\bar p$-annihilation into hadrons \cite{NPCS24} and decays of heavy quarkonia \cite{Tomsk}. Physicists learn the so-called "pure" annihilation, equal to the difference of topological cross sections $\sigma _n (p\overline p)$ and $\sigma _n (pp)$ at the same energy, to analyze MD in $p\bar p$ annihilation
\begin{equation}
	\label{18}
	\Delta \sigma _n (p\bar p-pp)=\sigma _n (p\bar p)- \sigma _n (pp).
\end{equation}

This difference exhibits a two-humped structure (including maximal value at $n$ = 6), which is explained in GDM by the inclusion of three intermediate quark topologies \cite{Ann,Ann1}. The two main topologies are formed only from valence $q\bar q$-pairs with the formation of three neutral pions or two charged and one neutral. The third topology is formed with the participation of quarks arising from the decay of the gluon to $g\to q\bar q$-pair. This topology is responsible for the high multiplicity. 

Superposition of these topologists \cite{NPCS24} describes  well data \cite{Rush} and gives the qualitative explanation of the strange phenomenon: the appearance of the sharp peak for creation of three pion pairs (3$\pi ^+\pi ^-$) in $e^+e^-$ annihilation at the threshold of $p\overline p$ pair formation, close to 1.9 GeV \cite{Novos}.

The choice of intermediate quark topologies is stipulated by experimental data, which indicate the absence of leading particles and the formation of three pion clusters (jets) \cite{Rush}.

Decays of heavy quarkonia are described by the convolution of three gluon jets and their hadronization \cite{Tomsk}. Also it is still not possible to explain the peaks of topological cross sections of $e^+e^-$annihilation in the region close to the threshold of a proton-antiproton pair ($\sim $1.8 GeV) formation \cite{Novos}. It is possible that at the initial stage a corresponding quark topology with a gluon component is formed.

\section{GDM and neutral pions}
While working on the Thermalization project, physicists from BITF (Kyiv, Ukraine) Mark Gorenshtein and Victor Begun \cite{BEC,BEC1} suggested that we search in the high multiplicity region for a signal of the formation of a pion (Bose-Einstein) condensate. In the energy range of U-70, pions, as the lightest hadrons, are produced abundantly. They are bosons and can form a pion condensate if their energy is low enough.

As the number of secondary particles increases, their average energy decreases, and, possibly, they gradually fall out into a condensate. In \cite{BEC,BEC1} it was shown that a signal of a pion condensate formation would be a sharp increase in the scaled variance $\omega ^0$ for the total multiplicity $n_{\textnormal {tot}}$ = $n_{ch}$ + $n_0$ ($n_0$ is the multiplicity of neutral pions) as a function of $n_{\textnormal {tot}}$
\begin{equation}
	\label{eq18}
	\omega ^0 (n_{\textnormal {tot}})= \frac {D_2}{\overline n_0} = \frac {\overline {n_0^2}-{\overline n_0}^2}{\overline n_0}.
\end{equation}

Charged particles are registered by the vertex detector. At that, corrections were made for the acceptance of the vertex detector, the registration efficiency, and the operation of a scintillation hodoscope, which suppresses low-multiplicity events. The systematic errors in determining the topological cross-sections were eliminated by ``stitching'' our data with measurements at the Mirabelle bubble chamber in the region of a trigger threshold value 8.

Monte Carlo simulation of our setup showed that almost 95~$\% $ of all $\gamma $-quanta entering the electromagnetic calorimeter are decay products of neutral pions. The lower threshold for their registration is about 100 MeV. At the same time, only 37 $\% $ photons from all $\pi ^0$-mesons enter the calorimeter. For half of them, both photons are registered, for the other half -- only one.

Simulations with using of several generators showed that the scaled variance of the neutral pion multiplicity remains constant with increasing total one. A linear dependence of average multiplicity of $\pi ^0$-mesons versus the number of \cite{BE} photons entering the calorimeter was also found.

The availability of a strong correlation between the multiplicities of neutral pions and the $\gamma $-quanta registered in the calorimeter allows, without reconstructing neutral pions, to measure the scaled variance for photons as a function of $n_{\textnormal {tot}} ^{(\gamma )}= n_{ch} +n_{\gamma }$, in order to make a conclusion about the behavior of $\omega ^0$ for $\pi ^0$-mesons.

Despite this possibility, our Collaboration has reconstructed the MD of $\pi ^0$ mesons. We did not use the event-by-event method, since it is impossible to determine the number of neutrals in each individual event. An original algorithm was proposed and implemented to reconstruct the number of events with a given multiplicity of neutral pions.

This algorithm is based on Monte Carlo simulation of the setup operation and is implemented in two stages. It is performed at each multiplicity of charged particles, starting with $n_{ch}$ = 4. In the first stage, the number of simulated events with $j$ $\pi ^0$-mesons $N_{ev}(j)$ is determined. Among them there will be $N_{ev}(i,j)$ events, when $i$ decay photons enter the calorimeter. Their ratios define the weighting coefficients $c_{ij}$~=~$ N_{ev}(i,j)/\sum _{i} N_{ev}(i,j)$. The number of events with $j$ $\pi ^0$-mesons is $N_{ev}(j)$ = $\sum _{i} N_{ev}(i,j)$, and the coefficients $c_{ij}$ determine what fraction of them are events with $i$ photons that hit the calorimeter.

At the second stage, the inverse problem is solved. Using the weighting coefficients $c_{ij}$ and the experimentally determined number of events with $i$ photons registered in the calorimeter ($N_{ev}(i)$), the number of events with $j$ pions $N_{ev}(j)$ is restored. It is carried out by multiplying $c_{ij}$ by $N_{ev}(i)$ and summing of these products in the cell of this table corresponding to the sought multiplicity of neutral pions. Note that this table allows us to determine the distribution by $n_{\textnormal {tot}}$, and other characteristics of MD's for charged and neutral particles. 

\begin{figure}[ht] 
	\leavevmode
	\centering
	\includegraphics[angle=0, width=0.42\textwidth]{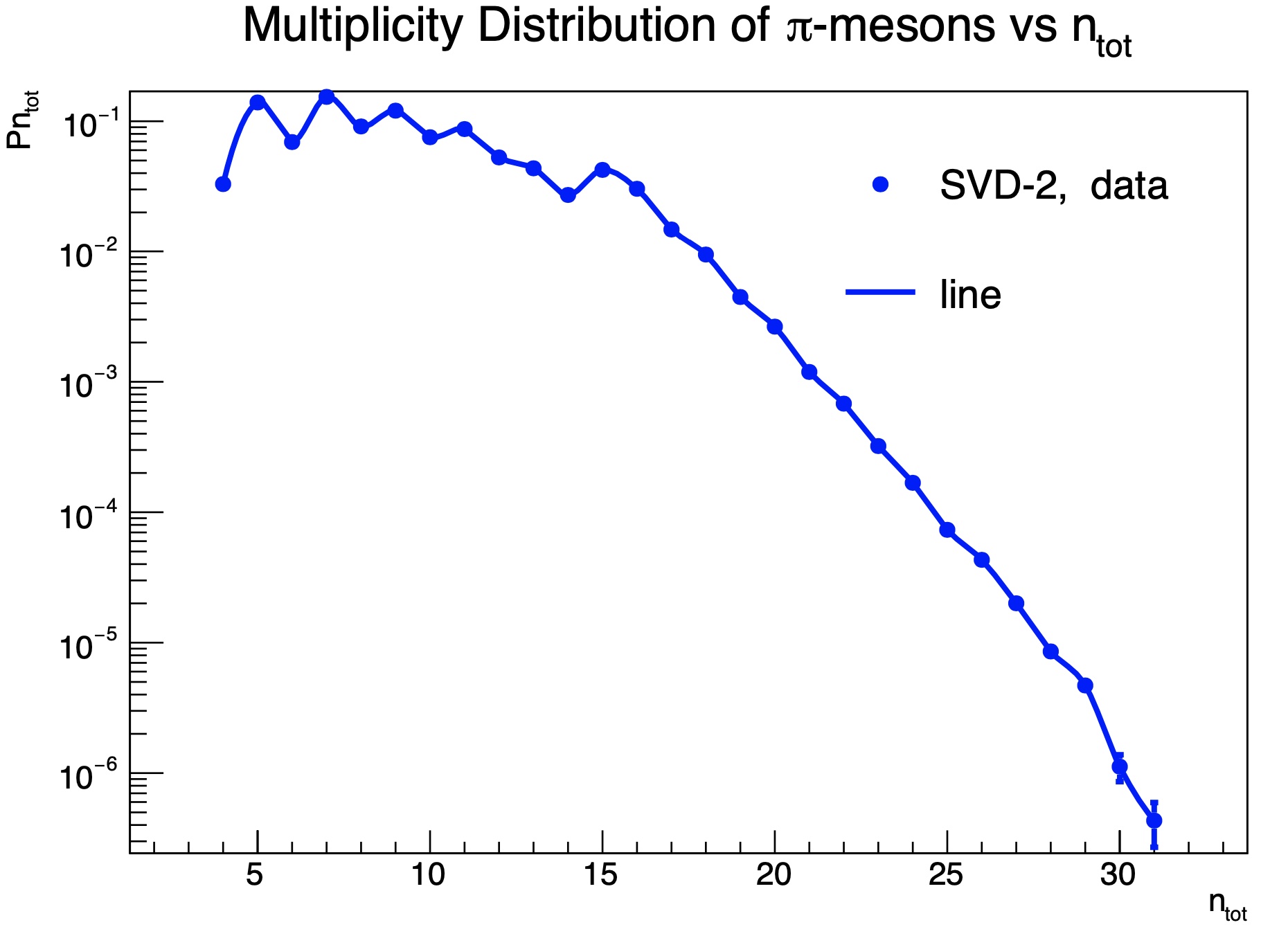}
	\includegraphics[angle=0, width=0.44\textwidth]{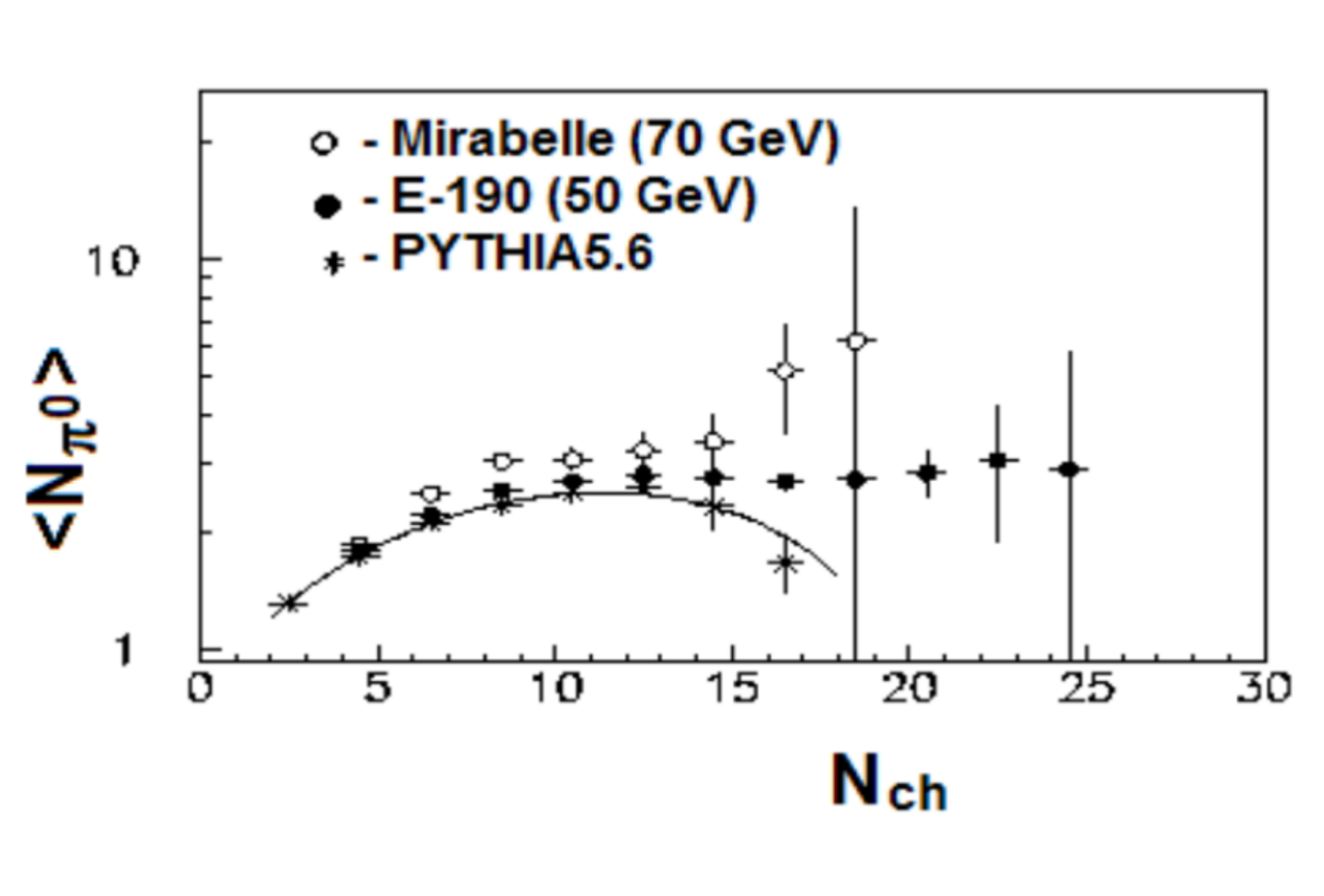}
	\caption{Experimental MD $Pn_{\textnormal {tot}}$ on the total pion multiplicity (left) and the average number of neutral mesons $<N_{\pi ^{0}}>$ on charged multiplicity $N_{ch}$ (right)
    \cite{BE}.} 
	\label{fig3}
\end{figure}

In Fig. \ref{fig3} (left), an experimental distribution by the total multiplicity of pions, $n_{\textnormal {tot}}$ is presented. It discovers unusual jumping behavior at low multiplicity. It can be explained by an even number of charged particles and any multiplicity of neutrals.

The dependence of the average number of $\pi ^0$-mesons $<N_{\pi ^{0}}>$ on $N_{ch}$, shown in Fig. \ref{fig3} on the right, shows a weak growth, underestimating their values. This behavior is due to the loss of some photons from the decay of $\pi ^0$-mesons at registration them by our calorimeter. It is stipulated by the existence $\approx$ 100 MeV-threshold of photon registration. For comparison, Mirabelle data and PYTHIA5.6 predictions are presented. They agree at low multiplicity.  At the same time, the whole interval for measured of topological cross-sections covers seven orders of magnitude.

\begin{figure}[ht] 
	\leavevmode
	\centering
	\includegraphics[angle=0, width=0.46\textwidth]{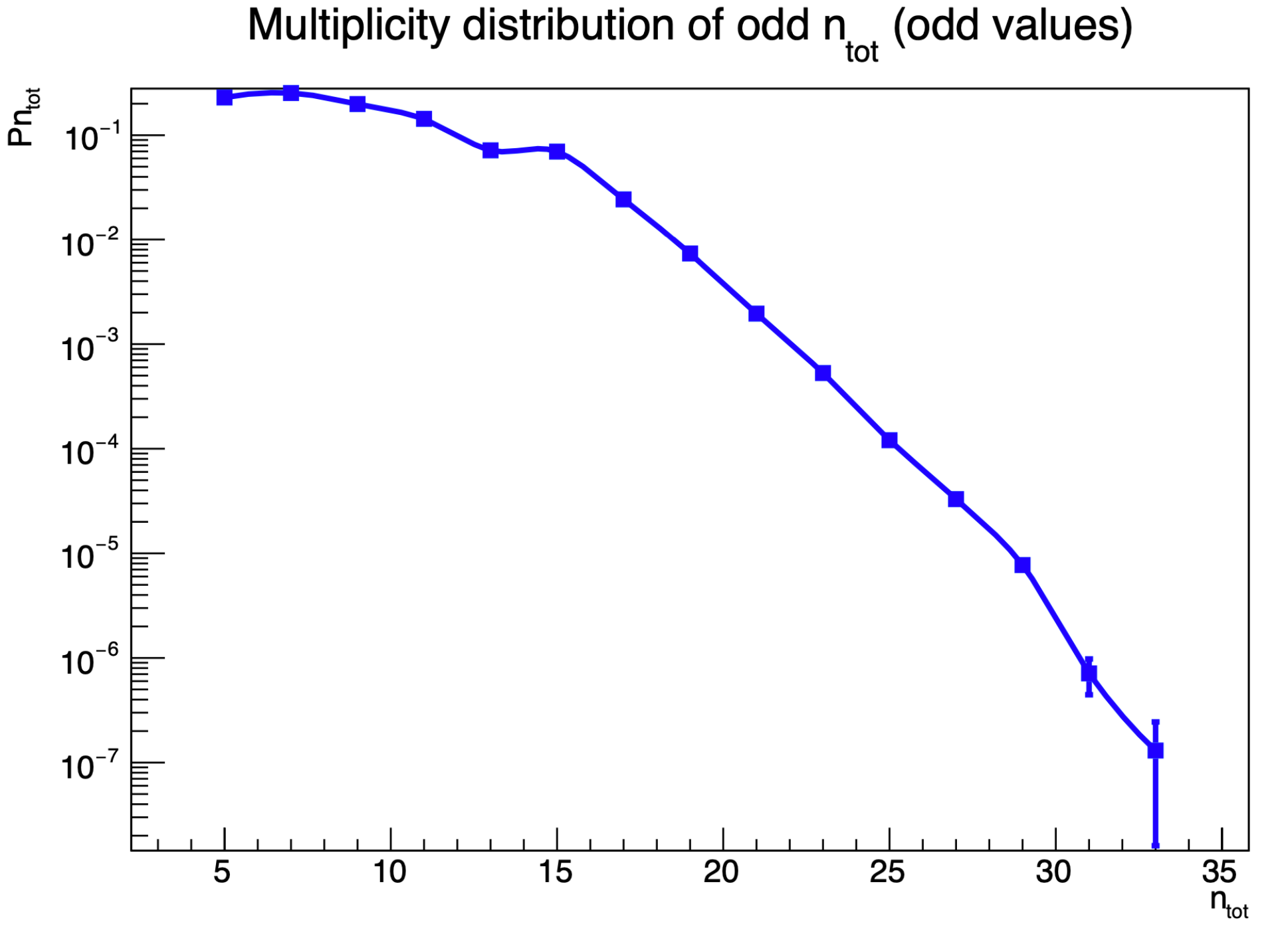}
	\includegraphics[angle=0, width=0.46\textwidth]{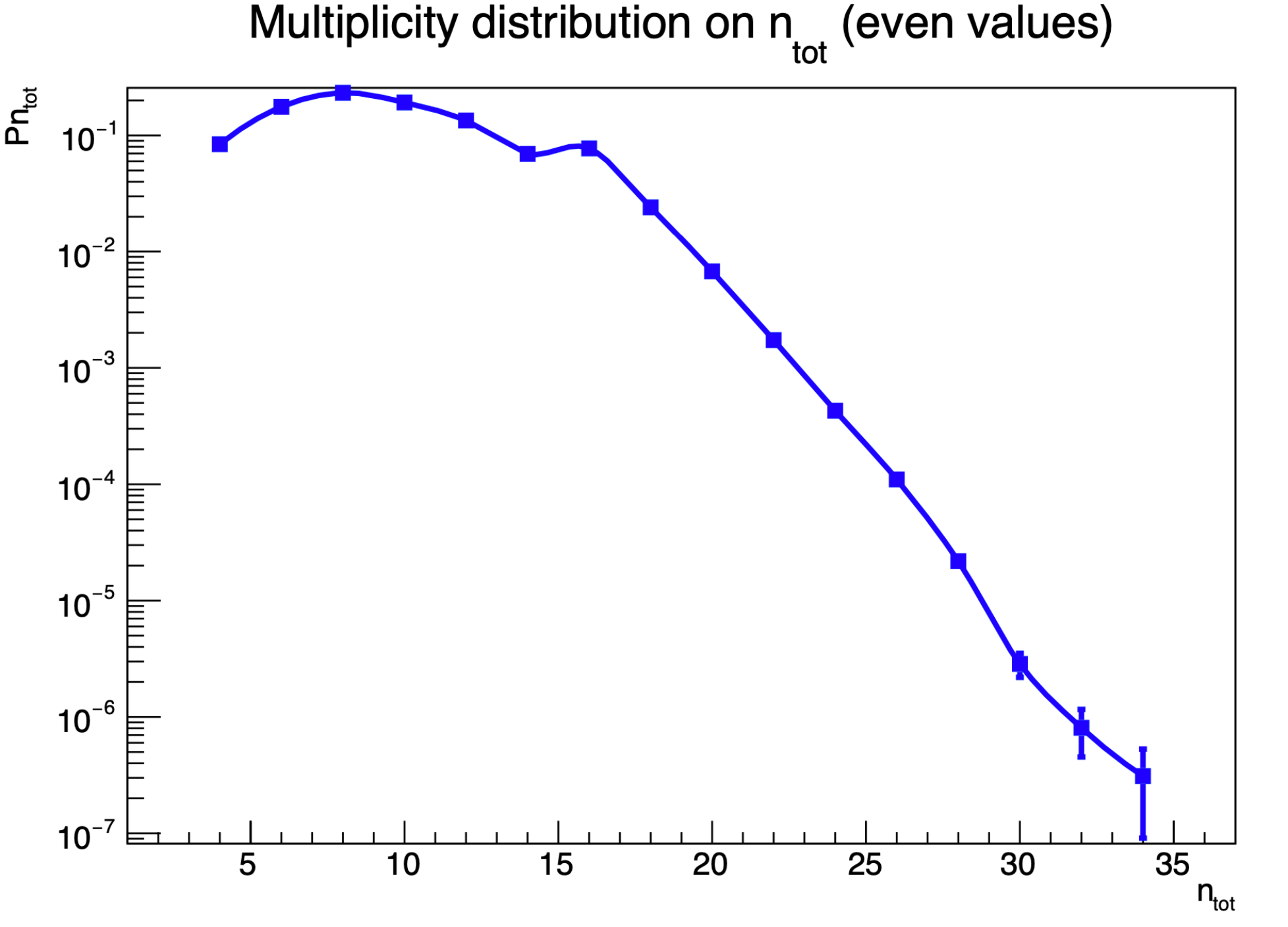}
	\caption{MD's on odd (left) and even numbers of neutral pions (right)\cite{BE}.} 
	\label{fig4}
\end{figure}

In fig. \ref{fig4} MD's for odd and even values of the total pion multiplicity are shown. It is seen that the jumping behavior at low multiplicity disappears. A curved line connecting experimental points is almost smooth (\ref{fig4}). 

The reconstruction of MD's on $\pi ^0$ mesons allowed us to verify the assumption \cite{BEC, BEC1} about the formation of a pion condensate. Experimental data confirmed the growth of $\omega ^0$ at the level of seven standard deviations in the high multiplicity region ($n_{ch} > $ 18 or $n_{\textnormal {tot}} > $ 26) \cite{BE} in comparison to the Monte Carlo simulation predictions. A sharp increase in the scaled variance $\omega ^0$ is detected in direct measurements from $n_{\textnormal {tot}}^{(\gamma )}$ for photons, and indirectly from $n_{\textnormal {tot}}$ for neutral pions \cite{BE}.

Now we will apply GDM to describe MD's for $\pi ^0$-mesons. Let us imagine that our setup registers only neutral pions. According to GDM, the sources of secondary hadrons are active gluons. Some of them participate in the production of charged pions, and the rest produce $\pi ^0$-mesons.

Since active gluons appear randomly, depending on the collision centrality we will describe them by a Poisson distribution. For the fragmentation of gluons into hadrons we use the universal Bernoulli distribution (\ref{eq1}) with hadronization parameters $\overline n^h_0$ and $N_0$ (the average and maximum possible multiplicity of neutral pions during gluon hadronization, respectively).

The Poisson distribution is $P_m = e^{-\overline m}\overline m^m /m!$,
where $\overline m$ is the average gluon multiplicity. The convolution of two stages under the soft de-coloration condition is written as

\begin{equation}
	\label{eq19}
	P_n(s) = \alpha \sum _{m=1}^{MG} \frac{\overline m ^m e^{-\overline m}}{m!}
	C_{mN_0}^{n} \left(\frac{{\overline n}^h_0}{N_0} \right )^n \left(1-\frac{{\overline n}^h_0}{N_0} \right )^{mN_0-n},
\end{equation}
where $MG$ is the maximum number of active gluons, $\alpha $ is the normalization factor.

\begin{figure}[ht] 
	\leavevmode
	\centering
	\includegraphics[angle=0, width=0.47\textwidth]{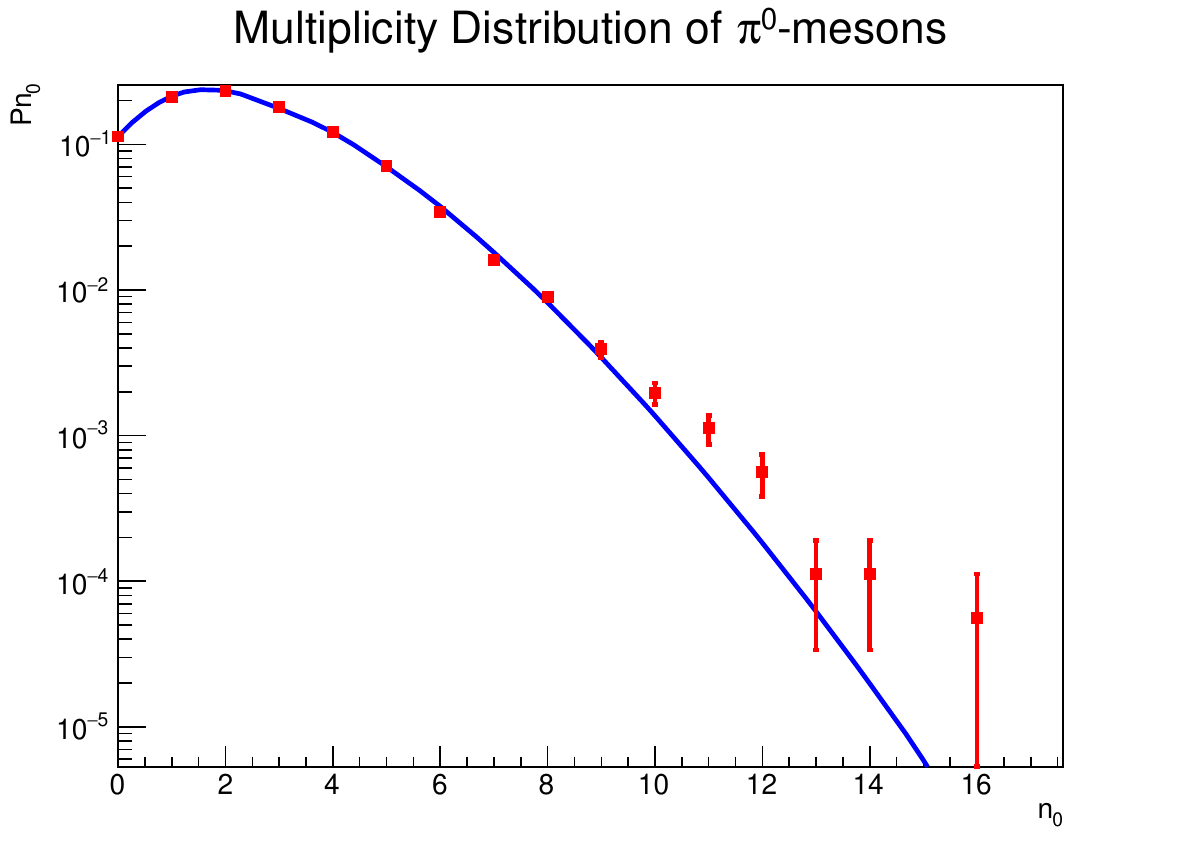}
	\includegraphics[angle=0, width=0.47\textwidth]{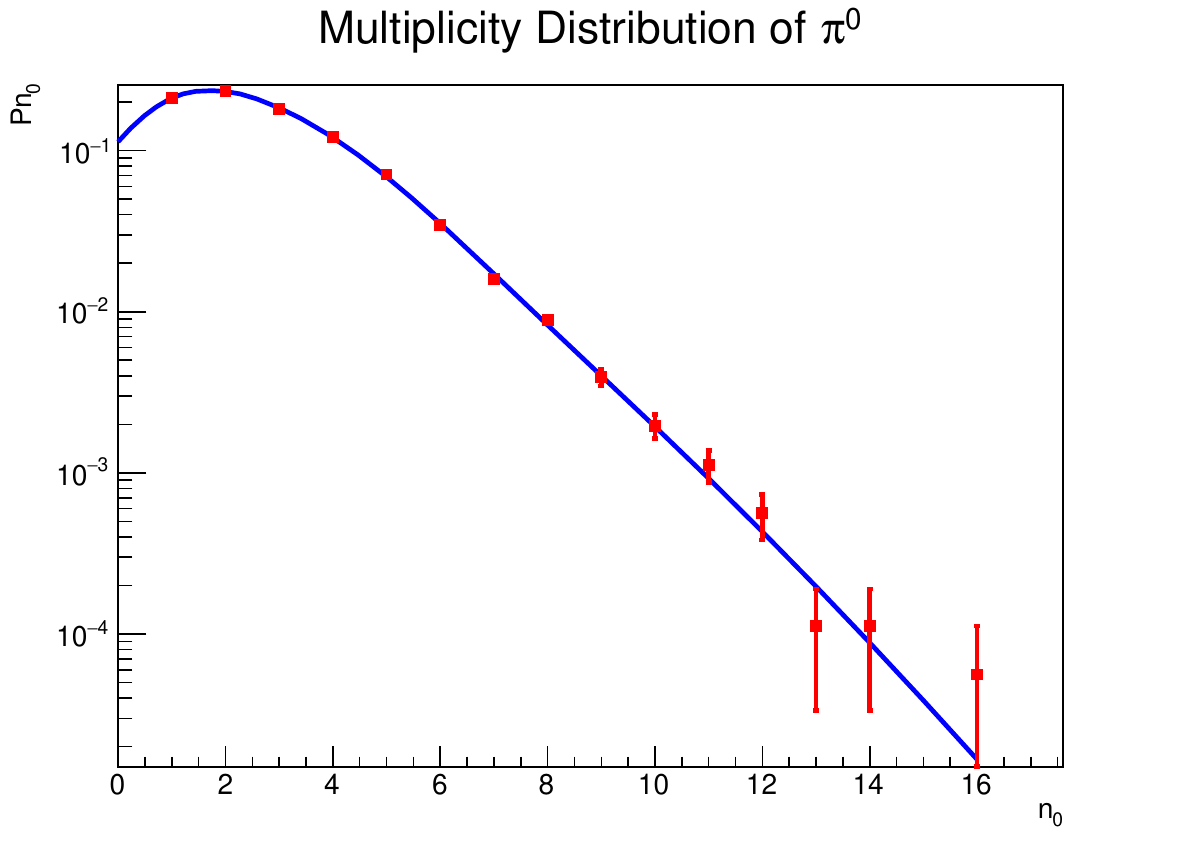}
	\caption{GDM description (blue line) of the MD for $\pi ^0$ mesons in the absence of gluon fission (left) and with it (solid line, right). Experimental data (shown $\blacksquare) $ are taken from \cite{BE}.} 
	\label{fig2}
\end{figure}

After comparing of the expression (\ref{eq19}) with the \cite{BE} data, following values of the model parameters have been obtained:
$\alpha $ = 1.06 $\pm $ 0.02, $\overline m$ = 2.39 $\pm $ 0.18, $N_0$ = 2, $\overline n^h_0$ = 0.99 $\pm $ 0.06, and $MG$ = 2. At this, $\chi ^2$~=~31.47 for $NDF$ = 12 turned out not so good. In Fig. \ref{fig2}, on the left, MD in GDM is shown. Obviously, GDM without gluon fission significantly underestimates MD at high multiplicity region. 

The similar situation arose at the description of MD for charged particles in the high multiplicity region, that required taking into account the gluon fission. Therefore, we write MD as a superposition of two contributions, without and with gluon fission:

$$
P_n(s) = \alpha _{1}\sum _{m=1}^{MG} \frac{\overline m_{1} ^m e^{-\overline m_{1}}}{m!}
C_{mN_0}^{n} \left(\frac{{\overline n}^h_0}{N_0} \right )^n \left(1-\frac{{\overline n}^h_0}{N_0} \right )^{mN_0-n} +
$$
\begin{equation}
	\label{eq20}	
	+ \alpha _{2}\sum _{m=1}^{MG} \frac{\overline m_{2} ^m e^{-\overline m_{2}}}{m!}
	C_{2mN_0}^{n} \left(\frac{{\overline n}^h_0}{N_0} \right )^n \left(1-\frac{{\overline n}^h_0}{N_0} \right )^{2mN_0-n},
\end{equation} 
where $\overline m_{1}$ and $\overline m_{2}$ are the average gluon multiplicities for events without and with their fission, respectively, $\alpha _{1}$ and $\alpha _{2}$ are normalization coefficients.

The description of experimental data by expression (\ref{eq19}) is presented in Fig. \ref{fig2}, on the right. The value $\chi ^2$ = 8.46 at $NDF$ = 10 has been improved significantly. The fitting parameters are as follows:
$\alpha _{1}$ = 1.02 $\pm $ 0.10, $\alpha _{2}$~=~1.18~$\pm $~0.37, $\overline m_{1}$ = 1.25 $\pm $ 0.12, $\overline m_{2}$ = 0.28 $\pm $ 0.14$, N_0$ = 7.55 $\pm $ 2.57,  $\overline n^h_0$ = 1.42 $\pm $ 0.06.

Our project "Thermalization" has been stopped in 2015, as the laboratory was involved in preparing an experiment at the future NICA collider. For it, we proposed a physical program to search for pion condensate and study the increased yield of soft photons \cite{GDM3}.

Thus, inclusion fission gluons in the scheme significantly improves $\chi ^2$. They are responsible for the occurrence of high multiplicity events for neutral pions, also as for charged particles \cite{GDM3}. This result can be used to describe the MD of other mesons and even baryons in high-energy events.

\section{Conclusion}
In this paper we have shown that GDM can describe MD for different processes by a two-stage scheme. They unite $e^+e^-$ and $p\bar p$ annihilation, proton collisions, and decays of heavy quarkonia. The phenomenological scheme of hadronization turned out to be universal for all these interactions. It confirms the change of hadronization mechanisms from fragmentation in $e^+e^-$-annihilation (vacuum) to recombination ($qg$-medium) hadron collisions.

GDM explains the production of secondary particles by active gluons, leaving valence quarks as observers in the leading particles (protons), which emphasizes the important role of the gluon component of the proton. Moreover, the active gluon fission is also the source of high multiplicity. These and other results obtained in GDM is causing increased interest to MP.

The authors express their sincere gratitude to SVD-2 Collaboration for their significant contribution to MP study of MD and the implementation of main tasks at the Thermalization project. We also hope that the study of multiparticle production will be continued in the SPD project, since it will have significantly greater capabilities for this.
\begin{center}
	
\end{center}

\end{document}